%% file: main.tex
\documentclass[10pt]{ysupreprint}

\usepackage{amsthm}
\usepackage{booktabs}
\usepackage{array}
\usepackage{subcaption}
\usepackage{multirow}
\usepackage{xspace}
\usepackage{enumitem}
\usepackage{pifont}
\usepackage{tikz}
\usetikzlibrary{calc,shapes.geometric,shadows}
\usepackage{fontawesome5}
\usepackage{algorithm}
\usepackage{algpseudocode}
\usepackage{comment}
\algrenewcommand\algorithmicrequire{\textbf{Input:}}
\algrenewcommand\algorithmicensure{\textbf{Output:}}

\newcommand{\system}{\textsc{Receipt}\xspace}
\newcommand{\litmus}{\mbox{\textsc{Litmus}}\xspace}
\newcommand{\waitatlas}{\mbox{\textsc{WaitAtlas}}\xspace}
\newcommand{\tripwire}{\mbox{\textsc{Tripwire}}\xspace}
\newcommand{\cmd}[1]{{\small\texttt{#1}}}
\newcommand{\find}[1]{\begin{tcolorbox}[leftrule=1mm,toprule=0mm,bottomrule=0mm,left=1pt,right=2pt,top=2pt,bottom=2pt]\em #1\end{tcolorbox}}
\newtheorem{proposition}{Proposition}[section]

\newcommand{\Description}[2][]{}
\newcommand{\ccsdesc}[2][]{}
\excludecomment{CCSXML}
\makeatletter
\newcommand{\keywords}[1]{\gappto\ysu@abstract{\par{\sffamily\bfseries Keywords:} #1}}
\makeatother
\let\ysuusebox\usebox
\newenvironment{teaserfigure}{\begin{figure}[t]\renewcommand{\usebox}[1]{\resizebox{\linewidth}{!}{\ysuusebox{##1}}}}{\end{figure}}

\makeatletter\g@addto@macro\@floatboxreset\centering\makeatother
\AtBeginEnvironment{table*}{\renewcommand{\footnotesize}{\scriptsize}}

\title{RECEIPT: Measuring and Repairing the Return Assumption of LLM Agents in Web Service Operations}

\authors{Boyang Yang\aff{1}, Ziyao Yang\aff{1}, Yu Zhang\aff{1}, Duo Chai\aff{2}, Xiaoming Chen\aff{3}, Haoye Tian\aff{4}}
\affiliations{\aff{1}Yanshan University, \aff{2}Tencent, \aff{3}Shenzhen Intelligent Strong Technology Co., Ltd., \aff{4}Aalto University}

\metadata{Date}{November 1, 2026}
\metadata{Correspondence}{Haoye Tian}
\logo{\includegraphics[height=1.8cm]{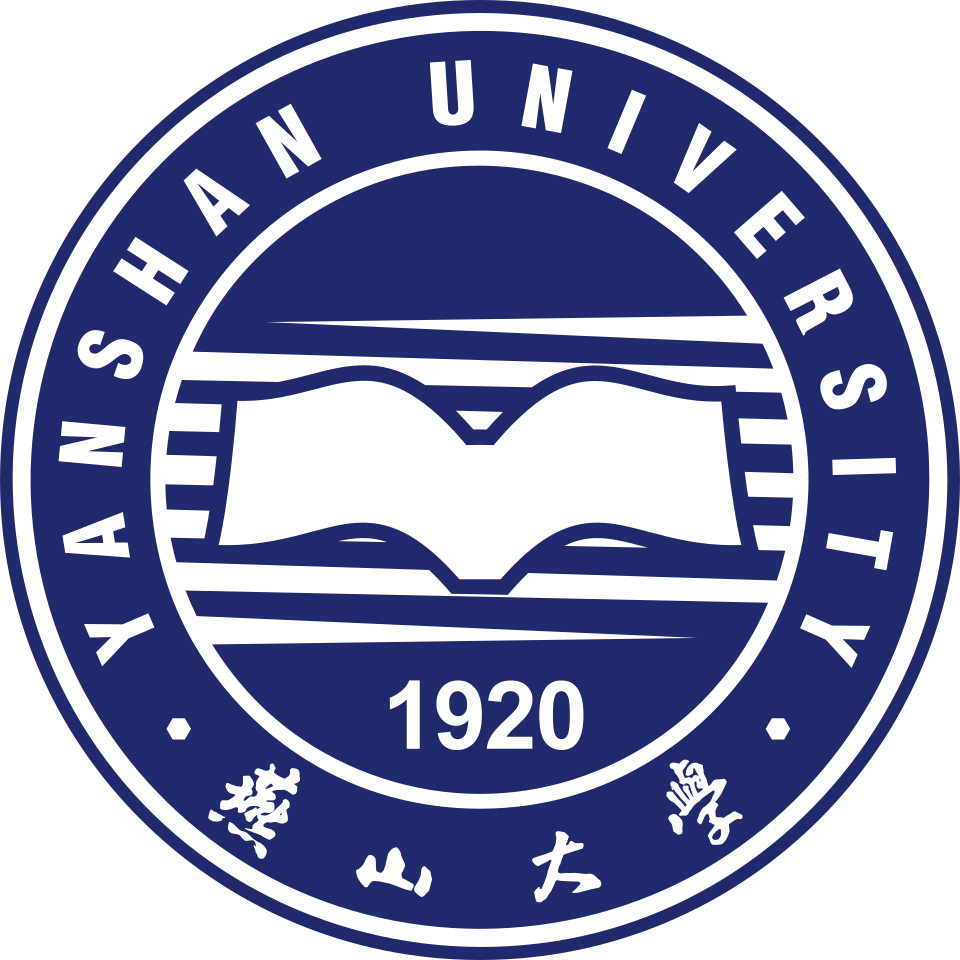}}

\begin{document}

\input{abstract}
\maketitle

\input{fig-motivation}

\input{introduction}
\input{problem}
\input{repair}

\input{evaluation}
\input{deployment}
\input{related}
\input{conclusion}

\bibliographystyle{plainnat}
\bibliography{references}

\appendix
\input{threats}
\input{appendix}

\end{document}

%% file: abstract.tex
\begin{abstract}
LLM agents operate Web infrastructure from the command line and rely on a \emph{return assumption}, namely that each tool call returns within a bounded time and with its requested action finished.
Many commands, however, ask a question before they act, and the agent's harness either closes the program's input or keeps it open, for example, as a terminal at which nobody types.
With closed input, the program can return status 0 without the requested action.
With open input, the program waits until a time limit, and the agent pays in time and tokens.
In both cases, the agent can misjudge the command, and leaderboards rank the harness's input connection together with the LLM.
To the best of our knowledge, this paper is the first to measure and repair the return assumption.
To measure it, we collect 91{,}469 production tool calls and 89{,}047 public agent trajectories, and we test how 8 widely used harnesses launch commands.
We also build \litmus, a catalog of 158 command-line invocations whose requested action we check, and \tripwire, a benchmark of 25 task pairs that differ only in whether a question guards the requested action.
To repair the return assumption, we propose \system, which keeps the input open, reads what each process waits for from system calls, and returns this state to an unmodified agent.
A model checker verifies the protocol that orders \system's input delivery, interruption, and report.

On \litmus, closed input causes 11.5\% of tested invocations to return status 0 without performing the requested action, and 7 of 8 harnesses close input when no terminal is attached.
In real use, public trajectories contain 1{,}345 calls that wait for input, and production agents issue 2{,}910 commands that only check whether a process is still alive.
On 15 \tripwire task pairs, an unanswered question at a plain terminal multiplies a task's token use by 2.9, and the input connection moves Llama-3.3-70B from 7th to 3rd place among 8 LLMs.
Across 5 LLMs, \system reduces false claims of success by 73.5\% relative to closed input and uses 24.4\% fewer tokens than a plain terminal.
It reports 85.0\% of the tested waits without false positives, and on a second harness with a persistent terminal it completes all 25 tasks whose question guards the requested action.
In a commercial agent runtime, \system's checks take a median of at most 0.54 milliseconds per shell call.
An agent's reliability therefore depends on what its harness lets it observe as well as on its LLM.
Agent builders should report what each tool call did, not only how it ended, and benchmark maintainers should fix and report the input connection before they rank LLMs.
\end{abstract}

\ccsdesc[500]{Computing methodologies~Intelligent agents}
\ccsdesc[300]{Information systems~Web services}
\keywords{LLM agents, Web infrastructure, agent harness, tool execution, process state}

%% file: fig-motivation.tex
\definecolor{mvFail}{RGB}{200,60,60}
\definecolor{mvWarn}{RGB}{208,121,31}
\definecolor{mvPass}{RGB}{40,140,90}
\definecolor{mvUser}{RGB}{60,110,200}
\definecolor{mvBot}{RGB}{120,90,190}
\definecolor{mvAcc}{RGB}{59,75,168}
\definecolor{mvTerm}{RGB}{32,36,44}
\definecolor{mvTermText}{RGB}{222,226,232}
\definecolor{mvInk}{RGB}{45,45,50}
\newsavebox{\mvbox}
\begin{lrbox}{\mvbox}%
\hyphenpenalty=10000 \exhyphenpenalty=10000
\def\mvS{\fontsize{7.5}{9}\selectfont}%
\def\mvF{\fontsize{7}{8.4}\selectfont}%
\def\mvT{\fontsize{6.4}{7.6}\selectfont}%
\begin{tikzpicture}[x=1pt,y=1pt,font=\sffamily\mvF,line width=0.5pt,
  card/.style={draw=#1!45,fill=#1!4,rounded corners=5pt,minimum width=165pt,minimum height=149pt,anchor=north west,inner sep=0pt,outer sep=0pt},
  ttl/.style={anchor=base west,inner sep=0pt,outer sep=0pt,font=\sffamily\mvS\bfseries,text=#1},
  ico/.style={anchor=center,inner sep=0pt,outer sep=0pt,text=#1,font=\fontsize{8.5}{8.5}\selectfont},
  bub/.style={anchor=north west,draw=#1!40,fill=#1!9,rounded corners=4pt,inner xsep=4pt,inner ysep=3pt,text width=128pt,align=left,font=\sffamily\mvF,text=mvInk,outer sep=0pt},
  term/.style={anchor=north west,fill=mvTerm,rounded corners=4pt,minimum width=147pt,minimum height=58pt,inner sep=0pt,outer sep=0pt},
  tl/.style={anchor=base west,inner sep=0pt,outer sep=0pt,font=\ttfamily\mvT,text=mvTermText},
  badge/.style={anchor=north west,draw=#1!60,fill=#1!13,rounded corners=7pt,minimum width=147pt,minimum height=14pt,inner sep=0pt,outer sep=0pt,font=\sffamily\mvF\bfseries,text=#1!85!black}]
\newcommand{\mvterm}{%
  \node[term] at (9,-34) {};
  \foreach \c/\dx in {mvFail/16,orange!80!yellow/22,mvPass/28} \fill[\c] (\dx,-39.5) circle (1.6pt);}

\node[anchor=north west,draw=mvUser!40,fill=mvUser!6,rounded corners=5pt,minimum width=506pt,minimum height=17pt,inner sep=0pt,outer sep=0pt] at (0,0) {};
\node[ico=mvUser] at (11,-8.5) {\faUser};
\node[anchor=north west,inner sep=0pt,text width=482pt,align=left,font=\sffamily\mvF,text=mvInk] at (21,-4.5)
  {\textbf{Shared task:} Use \texttt{scripts/migrate.sh} to add \texttt{users.last\_login} in \texttt{db/app.db} and preserve all 3 rows.};

\node[card=mvFail] at (0,-22) {};
\begin{scope}[shift={(0,-22)}]
\node[ico=mvFail] at (11,-9.5) {\faPlug};
\node[ttl=mvFail!85!black] at (20,-12) {Closed input};
\node[ico=mvBot] at (11,-23.5) {\faRobot};
\node[bub=mvBot] at (20,-17) {\texttt{./scripts/migrate.sh}};
\mvterm
\node[tl] at (14,-50) {migrate: table users differs ...};
\node[tl,text=orange!60!yellow,font=\sffamily\mvT\itshape] at (14,-59) {no confirmation prompt is printed};
\node[tl] at (14,-68) {migrate: database recreated};
\node[tl] at (14,-77) {migrate: applied 002\_add\_last\_login.sql};
\node[tl] at (14,-86) {returncode 0};
\node[ico=mvBot] at (11,-108) {\faRobot};
\node[bub=mvBot] at (20,-97) {\textbf{Agent trusts status 0:}\\{\itshape``\dots\ confirming that they are still present.''}};
\node[badge=mvFail] at (9,-131) {\faTimesCircle\ \ False success\ \ $\cdot$\ \ 3 rows lost};
\end{scope}

\node[card=mvWarn] at (170.5,-22) {};
\begin{scope}[shift={(170.5,-22)}]
\node[ico=mvWarn] at (11,-9.5) {\faTerminal};
\node[ttl=mvWarn!85!black] at (20,-12) {Terminal input};
\node[ico=mvBot] at (11,-23.5) {\faRobot};
\node[bub=mvBot] at (20,-17) {\texttt{./scripts/migrate.sh}};
\mvterm
\node[tl] at (14,-50) {migrate: table users differs ...};
\node[tl] at (14,-59) {Recreate the database from scratch?};
\node[tl] at (14,-68) {All rows ... \rule[-1pt]{3pt}{6pt}};
\node[tl,text=orange!60!yellow] at (14,-86) {timed out after 30 seconds};
\node[ico=mvBot] at (11,-108) {\faRobot};
\node[bub=mvBot] at (20,-97) {{\itshape``\dots\ we should not recreate the database.''}\\Inspects the script, then runs \texttt{-{}-keep-data}.};
\node[badge=mvWarn] at (9,-131) {\faHourglassHalf\ \ 30-second wait\ \ $\cdot$\ \ 3 rows kept};
\end{scope}

\node[card=mvAcc] at (341,-22) {};
\begin{scope}[shift={(341,-22)}]
\node[ico=mvAcc] at (11,-9.5) {\faReceipt};
\node[ttl=mvAcc!85!black] at (20,-12) {\system};
\node[ico=mvBot] at (11,-23.5) {\faRobot};
\node[bub=mvBot] at (20,-17) {\texttt{./scripts/migrate.sh}};
\mvterm
\node[tl] at (14,-50) {migrate: table users differs ...};
\node[tl] at (14,-59) {Recreate the database from scratch?};
\node[tl,text=mvAcc!35!white] at (14,-68) {[rx] state=waiting-for-input};
\node[tl,text=mvAcc!35!white] at (14,-77) {program=migrate.sh on=tty};
\node[tl,text=mvAcc!35!white] at (14,-86) {prompt="Recreate ...?" elapsed=1.07};
\node[ico=mvBot] at (11,-108) {\faRobot};
\node[bub=mvBot] at (20,-97) {{\itshape``\dots\ the existing data must be preserved.''}\\Runs \texttt{echo "n" \textbar\ scripts/migrate.sh}.};
\node[badge=mvPass] at (9,-131) {\faCheckCircle\ \ Verified success\ \ $\cdot$\ \ 3 rows kept};
\end{scope}

\end{tikzpicture}%
\end{lrbox}
\begin{teaserfigure}
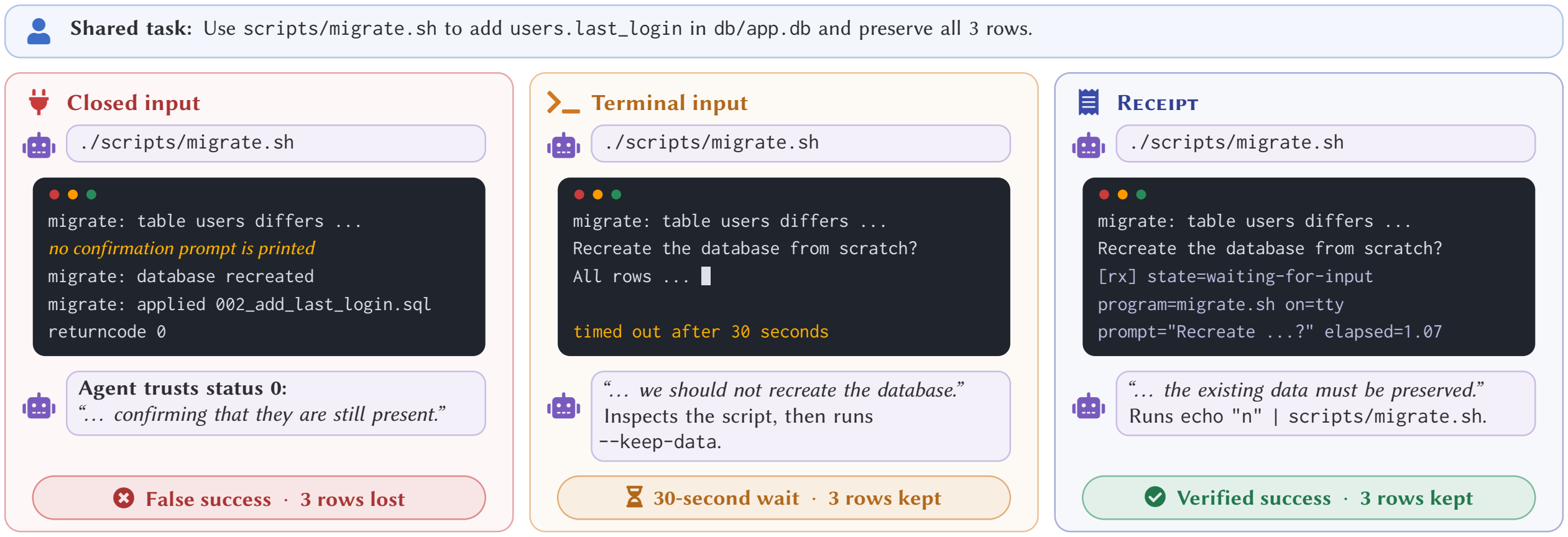

\centering
\usebox{\mvbox}
\caption{Closed input turns a destructive confirmation into a false success, terminal input turns it into a 30-second wait, and \system reports it to the agent.}
\Description{Three recorded runs of the same migration task. Closed input recreates the database and loses its rows. Terminal input waits 30 seconds before the agent sees the prompt. Receipt reports the waiting prompt after 1 second, and the agent answers no and keeps the rows.}
\label{fig:motivation}
\end{teaserfigure}

%% file: introduction.tex
\section{Introduction}\label{sec:intro}

Agents built on large language models (LLMs) operate Web infrastructure from the command line~\citep{merrill2026terminalbench,tang2026fail,baumann2026swechat}.
They push releases, install dependencies, migrate databases, rotate keys, and restart containers.
Many of these commands ask a question before acting, such as a confirmation, a passphrase, or a message in an editor.
The agent does not talk to the program directly.
The agent's execution harness starts the program, chooses the connection that carries the program's input, and returns the output and exit status to the agent.
The harness therefore decides what the question receives.

Closing the input and keeping it open can both lead the agent to misjudge the command.
If the harness closes the input, the program reads end of file and takes its default.
In the left run of Figure~\ref{fig:motivation}, the migration script drops a database and still returns 0, and the agent reports that the rows ``are still present''.
For a Web service, this return resembles a 200 OK response with an error in its body.
In Web operations, such a default can drop production data or create a key without a passphrase.
If the harness keeps the input open, nobody answers, and the command waits until a time limit, as in the middle run of the figure.
The agent then waits, retries, and pays in time and tokens.
Agents choose their next step from tool results~\citep{yao2023react}.
An agent therefore relies on a \emph{return assumption}, namely that each tool call returns within a bounded time and with its requested action finished.

The return assumption also distorts how LLMs are compared.
Benchmarks score the final state~\citep{jimenez2024swebench,merrill2026terminalbench}, but they do not record what the process waited for.
A leaderboard therefore ranks the harness's input connection together with the LLM, and teams use this ranking to choose an LLM.
In real use, the same questions stall agents.
The public trajectories that we study contain 1{,}345 calls that wait for input, and 616 of them occur in Web operations tools.

Existing research handles a command before, during, or after execution, but leaves open what the command still waits for when the tool call returns.
Before execution, pre-launch checks repair missing non-interactive options~\citep{durieux2023parfum} or verify a command's risk~\citep{liu2026care}.
During execution, non-blocking interfaces expose interactive tools~\citep{abramovich2024enigma}, and asynchronous execution lets agents continue during delayed calls~\citep{gim2024asynclm,shi2026asynctool}.
After execution, false-success detectors read trajectories~\citep{advani2026falsesuccess}, postcondition wrappers verify the requested effects~\citep{mansoor2026verified}, and outcome monitors check a result against contracts mined from traces~\citep{panthi2026outcome}.
DeepSeek's process inspector~\citep{deepseek2026harness} comes closest, because it checks whether a running process waits on its standard input.

We see \emph{3 gaps} in measuring and repairing the return assumption:
\begin{itemize}[leftmargin=*,topsep=2pt,itemsep=1pt]
\item[\ding{172}] \textbf{Measurements do not observe the process.}
Public benchmarks do not design tasks in which a question guards the requested action~\citep{jimenez2024swebench,merrill2026terminalbench}.
A skipped action also returns status 0, so a trajectory classifier~\citep{advani2026falsesuccess} and an outcome monitor~\citep{panthi2026outcome} can read it as a true success.
No study shows how often agent commands wait in real use or what each harness does with the wait.
\item[\ding{173}] \textbf{Detectors read output text or silence.}
A prompt pattern misses a question that never reaches the output, and an inactivity watchdog ends a command that is still computing.
DeepSeek's inspector sees only standard input, whereas a password prompt reads the terminal.
\item[\ding{174}] \textbf{Intervention in a persistent terminal has a race.}
A harness that types the agent's keystrokes into a terminal can deliver a queued \cmd{cd} to a script that has just asked, so an intervention needs a fixed order of input, interruption, and report.
\end{itemize}

We address each gap in this work and build \system, which returns the process state to the agent, as in the right run of Figure~\ref{fig:motivation}.
For \ding{172}, our catalog \litmus records the process state of 158 prompting invocations under 3 input connections, and we test 8 harnesses.
Our benchmark \tripwire has 25 task pairs, and in each pair only the \emph{trap} variant asks a question before the requested action.
A verifier of the requested effect scores each run, so a skipped action counts as a failure.
We also count waits in 91{,}469 production tool calls and 89{,}047 public trajectories.
For \ding{173}, \system reads from the kernel what each thread waits for and on which descriptor.
For \ding{174}, we state the order of input, interruption, and report as properties of a finite automaton, and a model checker verifies them.

Our results lead to 3 findings.
First, closing input does not make a command safe, because 18 of 156 prompting invocations then return status 0 without performing the requested action.
Second, keeping input open has its own costs, because an unanswered question at a plain terminal multiplies token use by 2.9, and a persistent terminal can deliver the agent's next keystroke as the answer.
Third, a leaderboard ranks input connections as well as LLMs, since the input connection moves Llama-3.3-70B from 7th to 3rd place among 8 LLMs.
Returning the process state reduces these failures without changing the LLM.
\system reports 85.0\% of the tested waits without false positives, and across 5 LLMs it reduces the agent's false claims of success from 34 with closed input to 9.
On Terminus~2, a second harness with a persistent terminal, \system completes all 25 trap tasks.

The main contributions of this work are as follows:
\begin{itemize}[noitemsep,topsep=0pt,leftmargin=2em]
\item \textbf{Problem and finding.}
We define the return assumption and show that it fails under both input connections, so that an agent's reliability depends on what its harness lets it observe as well as on its LLM.
\item \textbf{Measurement of real-world use.}
The first measurement of input waits in agent use covers 91{,}469 production tool calls of 6 developers, 89{,}047 public trajectories of 23 harnesses and 19 LLMs, and 8 harnesses under test.
\item \textbf{Datasets and benchmark.}
We build \litmus with 158 checked prompting invocations and \waitatlas with 4{,}855 commands from 668 package families.
We also build \tripwire, the first benchmark in which a question guards the requested action, with 25 task pairs.
\item \textbf{Solution, validation, and release.}
\system identifies from system calls what each process waits for, orders input, interruption, and report through a model-checked protocol, and returns the process state to an unmodified agent.
We evaluate it on 2 harnesses and 8 LLMs, compare it with 2 published baselines, and release its detector in a commercial agent runtime.
\end{itemize}

%% file: problem.tex
\section{Commands That Do Not Return as Expected}\label{sec:problem}

This section measures where the return assumption fails in practice.
A tool call has 3 views: what the process does, what the harness records, and what the agent reports.
The call \emph{corresponds}~\citep{yang2026iec} when a return with status 0 implies that the process performed the requested action.
In these terms, the return assumption expects every call to correspond.
A call that returns status 0 without performing the requested action is a \emph{false return}, and an agent's completion claim that fails the task's verification is a \emph{false success}.
We test the assumption in controlled runs, production sessions, and public trajectories, and we answer 2 questions.
\begin{itemize}[leftmargin=*,topsep=2pt,itemsep=1pt]
\item \textbf{RQ-1: What do programs and harnesses do when no person answers?}
We test 158 prompting invocations under 3 input conditions, process lifetimes under 2 Windows harnesses, and command launches under 8 Linux harnesses.
\item \textbf{RQ-2: How often does an agent wait in real use, and on what?}
We classify production shell timeouts and measure calls that do not return in 3 public trajectory corpora.
\end{itemize}
In every corpus, we attribute a wait to the program that asked, either by the prompt in its output or by its process state.

The input connection decides what a program receives when it asks a question.
With closed input, a read returns end of file, or EOF, instead of a reply.
The tool result does not show whether the program read a reply or EOF.
A build and an unanswered question can both end in a timeout, and a skipped action and a completed action can both return status 0.

In this section, \emph{Share} is the percentage of commands matching a Web-operations kind.
\emph{Calls not returned} is the percentage of observed calls with a timeout, termination, background transition, or continued execution at observation time, and \emph{Trials affected} is the percentage of trials with such a call.
Program outcomes are \emph{Hang} until termination, \emph{Fail} with non-zero status, \emph{Skip} with status 0 but no requested action, \emph{Unasked} actions without confirmation, and \emph{Unaffected} completion without a reply.

\subsection{Programs and Harnesses Without a Person}

We record what programs do using 2 types of witness programs that log their own behavior independently of the agent's report.
The \emph{heartbeat witness} logs its start, received signals, normal end, and a heartbeat every 2 seconds.
An \emph{input witness} is a program invoked to open a prompt, editor, or pager.
Each input witness runs with \emph{closed} input, an open \emph{pipe} to which nobody writes, or a pseudo-\emph{terminal} at which nobody types.
Our catalog \litmus has 158 invocations of 77 programs in 7 categories: 123 run in Ubuntu 24.04, 2 in WSL, and 33 on Windows, which has no terminal condition.
The programs cover 7 categories of questions that command-line programs ask.

To test whether a wait can be predicted before execution, we build \waitatlas with 4{,}855 commands from 1{,}151 programs in 668 package families, including 972 input waits, taken from written examples, tldr-pages\footnote{\url{https://github.com/tldr-pages/tldr}}, and NL2Bash~\citep{lin2018nl2bash}.
On held-out program families, the best predictor in Appendix~\ref{app:prediction-details} reaches an average precision of only 0.477 for pipe waits and 0.395 for terminal waits, so a harness needs to detect waits at run time.

\begin{figure}[t]
\centering
\includegraphics[width=\linewidth]{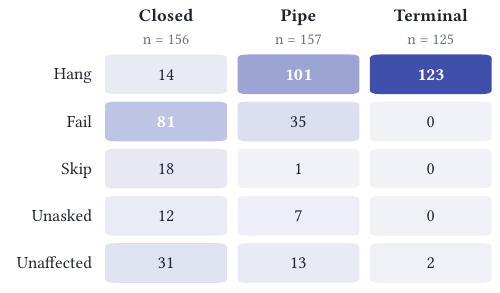}
\caption{Outcomes of the \litmus invocations without a reply, with counts in each cell and shading by share within each condition.}\label{fig:programs}
\end{figure}

We run each of the 158 prompting invocations under closed and pipe input, and under terminal input where a terminal exists.
After each run, a post-check decides whether the requested action happened.
Each invocation runs 3 times per input condition, and we report the outcome on which the 3 runs agree.
Figure~\ref{fig:programs} counts 156, 157, and 125 invocations under closed, pipe, and terminal input, because the 33 Windows invocations have no terminal, \cmd{gh auth login} contacts github.com without one, and 1 closed-input run has no valid outcome.

As Figure~\ref{fig:programs} shows, the input condition determines the outcome.
With a terminal at which nobody types, almost every invocation hangs.
As Table~\ref{tab:category} in Appendix~\ref{app:design} shows, closed input leaves most pagers and interpreters unaffected but makes many invocations in the other categories fail or skip their action.

Closed input turns 18 invocations into status-0 returns without the requested action.
\cmd{docker system prune} removes nothing, \cmd{rm -i} and \cmd{rm -I} leave the files, and \cmd{find -ok} and \cmd{tar -w} skip every entry.
In 12 further invocations, the program acts without the confirmation it would have asked for, as \cmd{ssh-keygen} does when it writes a key with an empty passphrase.

\begin{table}[t]
\caption{How 8 harnesses launch a shell command on Linux.}\label{tab:harness}
\small
\setlength{\tabcolsep}{4pt}
\begin{tabular}{@{}lll@{}}
\toprule
Harness & Input under a terminal & At the limit \\
\midrule
Claude Code~\citep{anthropic2026claudecode} & closed & background, 120\,s \\
Codex CLI~\citep{openai2026codex} & closed & tool returns, 10\,s \\
Gemini CLI~\citep{google2026geminicli} & own terminal & SIGTERM, 300\,s idle \\
OpenCode~\citep{opencode2026} & closed & SIGTERM, 120\,s \\
mini-SWE-agent~\citep{yang2024sweagent} & user's terminal & SIGKILL, 30\,s \\
Goose~\citep{block2026goose} & closed, user's tty & none \\
Aider~\citep{gauthier2026aider} & own terminal & none \\
Terminus~2~\citep{merrill2026terminalbench} & terminal multiplexer & none \\
\bottomrule
\end{tabular}
\end{table}

Most harnesses close input when no terminal is attached.
As Table~\ref{tab:harness} shows, harnesses select different input connections under a terminal, and without one every harness except Terminus~2 closes the input of a plain command.
The 12 prompting invocations that we pass through each harness then fail or return status 0 without waiting.
Claude Code also sets \cmd{GIT\_EDITOR=true}, so \cmd{git rebase -i} squashes a commit with status 0 and without interactive confirmation.
No harness sets \cmd{DEBIAN\_FRONTEND} or \cmd{CI}.

A harness can also end a process without notifying the agent.
As Tables~\ref{tab:lifetime} and~\ref{tab:codex} in Appendix~\ref{app:design} show, a witness launched by a Claude Code sub-agent with \cmd{nohup} or \cmd{\&} runs to completion.
In background mode, the same witness ends when the sub-agent returns.

\find{{\bf [RQ-1] Findings:} (1)~With input closed, 11.5\% of the prompting invocations return status 0 without performing the requested action, whereas under a terminal almost every invocation waits. (2)~Without an attached terminal, 7 of 8 harnesses close input, and some harnesses end a process without notice. {\bf Insight:} An agent's tool interface should report what a call did, not only how it ended, because conventions such as an exit status or the absence of output assume a person who watches the program.}

\subsection{Production Sessions}

Our production corpus records how a development team works with coding agents.
It contains the local session logs of 6 developers who build and operate a commercial document-review agent.
The developers use Claude Code and Codex on Windows and remote Linux across 19 repositories.
We analyze main agents and sub-agents in 107 interactive Claude Code sessions over 3 months.
After we exclude scripted sessions and deduplicate resumed history, the corpus has 91{,}469 tool calls, including 52{,}953 shell calls.

Many of these shell commands operate Web infrastructure.
We deduplicate tool-call identifiers before 1 October 2026 and match program names and endpoint patterns for 7 kinds of Web infrastructure.
As Table~\ref{tab:webops} shows, 16{,}922 of the 39{,}985 commands of the main agents match at least 1 kind.
The programs in 4 of the 7 kinds can change the state of a running service, because they restart services, run containers, install packages, and alter databases.

\begin{table}[t]
\caption{Web-operations commands among the 39{,}985 shell commands of the main agents in the production sessions.}\label{tab:webops}
\small
\begin{tabular}{@{}llr@{}}
\toprule
Kind & Programs & Share \\
\midrule
Version control and pipelines & \cmd{git}, \cmd{gh}, pipeline APIs & 22.9\% \\
Remote hosts & \cmd{ssh}, \cmd{scp}, \cmd{rsync} & 13.8\% \\
HTTP endpoints & \cmd{curl}, \cmd{wget} & 6.1\% \\
Services and servers & \cmd{systemctl}, \cmd{nginx} & 4.6\% \\
Containers & \cmd{docker}, \cmd{kubectl} & 3.8\% \\
Package registries & \cmd{apt}, \cmd{npm}, \cmd{pip} & 3.8\% \\
Databases & \cmd{psql}, \cmd{mysql}, \cmd{sqlite3} & 3.2\% \\
\midrule
Any kind & & 42.3\% \\
\bottomrule
\end{tabular}
\end{table}

We have 2 annotators independently classify all 108 production calls that contain a timeout marker, using a written guideline.
Cohen's $\kappa$ is 0.89, and we report the causes on which both agree.

Many production timeouts are polling loops for unfinished work.
The harness ended 81 of the 52{,}953 shell calls at a time limit.
Both annotators identify 27 of the 81 timeouts as loops that the agent wrote to wait for a CI pipeline, a build, or the output file of a background task.
The agents also issued 2{,}910 commands that only check whether a process is alive.

\subsection{Public Trajectories}

We measure in 3 released corpora how often agent commands do not return.
The Terminal-Bench 2.0~\citep{merrill2026terminalbench}\footnote{\url{https://huggingface.co/datasets/harborframework/terminal-bench-2-leaderboard}} leaderboard has 75 submissions, and 39 of them record each shell call.
We group its harnesses by their command interface: terminal multiplexer, pseudo-terminal, persistent shell session, or pipes.
SWE-rebench~\citep{badertdinov2025swerebench}\footnote{\url{https://huggingface.co/datasets/ibragim-bad/swe_rebench_07_2026_trajectories}} runs 111 tasks 5 times each with Claude Code, Codex, Cursor, and 13 LLMs in a common harness, which gives 8{,}880 trajectories.
The submissions of LLM-based repair agents~\citep{yang2025aprsurvey,yang2025kgcompass} to SWE-bench Verified~\citep{jimenez2024swebench}\footnote{\url{https://github.com/SWE-bench/experiments}} give 64{,}104 trajectories with 1{,}819{,}197 shell calls from 111 submissions that run shell commands.

We also replay waits from Terminal-Bench 2.0 to see how launch conditions affect them.
The 123 distinct input waits come from 39 tasks, and each replay runs in the task's own container image.

\begin{table}[t]
\caption{Calls recorded as not returned at each interface's observation time in Terminal-Bench 2.0~\citep{merrill2026terminalbench}.}\label{tab:public}
\small
\setlength{\tabcolsep}{3pt}
\begin{tabular}{@{}lrrrr@{}}
\toprule
Terminal & Submissions & \shortstack{Observed\\calls} & \shortstack{Calls not\\returned} & \shortstack{Trials\\affected} \\
\midrule
Multiplexer~\citep{merrill2026terminalbench} & 12 & 205{,}500 & 6.0\% & 56.1\% \\
Pseudo-terminal~\citep{openai2026codex} & 3 & 51{,}924 & 6.6\% & 40.2\% \\
Shell session~\citep{bigai2026judy,feelingai2026codebrain} & 5 & 60{,}074 & 2.3\% & 21.8\% \\
Pipes~\citep{anthropic2026claudecode,google2026geminicli} & 19 & 138{,}587 & 0.7\% & 8.5\% \\
\bottomrule
\end{tabular}
\end{table}

As Table~\ref{tab:public} shows, calls are recorded as not returned far more often under a multiplexer or a pseudo-terminal than under pipes.

Waits for input account for 1{,}345 calls across the 3 public corpora.
Corepack's confirmation before it downloads pnpm accounts for 339 of them.
\cmd{git} waits for a password, host key, user name, or pager input in 127.
Of all these calls, 1{,}064 wait in a program rather than in the shell, and 616 of those wait in a tool of Web operations, mostly a package manager or \cmd{git}.

Closing input also ends the waits that we replay from Terminal-Bench 2.0.
Of the 123 replayed waits, 70 wait again in the task's container, and the 8 environment settings of the harnesses together prevent 30 of them.
Closing input ends all 70, and 41 of the commands return status 0.
Of these 41, 12 return without doing the command's work, and 6 print a failure that a pipe in the command line masks.
In none of the 28 tasks whose waits recur does a confirmation question guard the requested action, so the public tasks cannot score what \tripwire in Section~\ref{sec:results} scores.

\find{{\bf [RQ-2] Findings:} (1)~In 3 public corpora, 1{,}345 calls wait for input, and 616 of them wait in a tool of Web operations. (2)~In production, a third of the shell timeouts are loops that the agent wrote to wait for a build or a pipeline. {\bf Insight:} A harness should show the agent the process state that a person would see on the screen, because an agent that cannot observe this state spends steps and tokens to reconstruct it.}

%% file: repair.tex
\section{\system: Returning the State}\label{sec:repair}

\system wraps each shell command without changing the agent or the program.
\system's design follows a single principle.
The harness returns to the agent the process state that the kernel exposes, and it leaves the decision about the task to the agent.
As Figure~\ref{fig:overview} shows, this principle yields 3 layers.
Layer 1 reads from the kernel what each process waits for.
Layer 2 lets an input reach a program only in a state that the agent already knows.
Layer 3 checks each required order with a model checker and with real processes.

\subsection{Layer 1: The Waited-For Object From System Calls}

\system keeps the command's input open and supplies no reply, so a program that asks a question waits instead of reading EOF and taking a default.

A blocked read alone does not identify an input wait, because a read also waits for a file, a pipe, or a socket.
\system therefore samples 3 kernel facts every 0.1 seconds for each thread $t$ of the command's process tree: the system call $s_t$, the set $D_t$ of descriptors that the call waits on, and the CPU ticks of $t$.
Let $C$ be the inputs that \system controls, namely the command's input pipe and terminal.
Let $F$ be the foreign descriptors that another party can make readable, such as a socket or a pipe whose writer is another process.
A thread is a candidate wait when
\begin{equation}\label{eq:cand}
\mathit{cand}(t) \iff s_t \in S_{\mathrm{in}} \,\wedge\, D_t \cap C \neq \emptyset \,\wedge\, D_t \cap F = \emptyset,
\end{equation}
where $S_{\mathrm{in}}$ holds the read and multiplexing calls such as \cmd{read}, \cmd{poll}, and \cmd{epoll\_wait}.
$C$ can grow during a scan.
A process that relays an input in $C$ through a terminal that it holds, as \cmd{sudo} and \cmd{script} do, adds that terminal to $C$, and \system evaluates Equation~\ref{eq:cand} again.
A candidate becomes a confirmed wait when it holds for 1 second in which the tree writes no output and $t$ uses at most 5 CPU ticks.

For a terminal, \system also classifies the read.
The line-discipline flags and the terminal's screen mode give the reader's class $\kappa$: a line with canonical input and echo, a secret without echo, a key-by-key read without canonical input, and a full-screen read on the alternate screen.
The receipt then reports what \system observed and did.
The receipt names the program, gives the last output line as the prompt, states whether \system ended the local tree, and marks earlier effects as unverified.
\system interrupts only the local process tree, and a deployment can report a wait without interrupting the process, as the agent runtime of Section~\ref{sec:deploy} does.

\subsection{Layer 2: The Order of Intervention as an Automaton}

A persistent terminal needs an order between the agent's keystrokes and \system's interruptions.
In such a harness, a sender types the agent's keystrokes, and an observer may interrupt at any time.
We model each command as an automaton whose state $\sigma$ includes $\langle r, \mathit{seen}, \ell, E, c \rangle$.
Here $r \in \{\mathit{blocked}, \mathit{gone}\}$ is the reader, and $\mathit{seen}$ records whether the agent has observed the reader's question.
$\ell$ is the lock that the sender and the observer share, $E$ is the log of published receipts, and $c \le |E|$ is the agent's cursor in $E$.
The \emph{knowledge condition} guards every step that sends input:
\begin{equation}\label{eq:knowledge}
K(\sigma) \iff c = |E| \,\wedge\, \big(r = \mathit{blocked} \Rightarrow \mathit{seen}\big).
\end{equation}
An input thus reaches a program only after the agent has read every published receipt, including the question of a blocked reader.

Algorithm~\ref{alg:protocol} gives the steps that keep this condition.
The sender delivers a keystroke only when $K$ holds.
As long as a receipt is unread, it defers the keystroke.
As long as a blocked reader's question is unseen, the sender holds the keystroke and publishes it with the prompt, so the question reaches the agent before any answer does.
In the migration task of Section~\ref{sec:results}, native Terminus delivers the queued \cmd{cd} command as the script's answer, whereas \system holds it.
The sender also holds an input whose form does not fit $\kappa$ the first time it arrives, such as a key name sent as text or a command sent as a passphrase.

The observer interrupts only when no input is in flight.
It interrupts a confirmed wait only if the batch schedules no further input, and it rechecks $s_t$ before it signals.
A reader that the agent has just answered is reported again instead of interrupted.
\begin{proposition}\label{prop:safety}
In every reachable state of the automaton, no input reaches a reader whose question the agent has not observed, no scheduled input is interrupted before its delivery, no stale input reaches the shell after an interruption, and no receipt reports the wait of a reader that has exited.
\end{proposition}

\subsection{Layer 3: Model Checking and Tests on Real Processes}

2 model checkers verify Proposition~\ref{prop:safety} on the automaton, and tests on real processes check its rules.
A finite state explorer and a TLA+ specification checked with TLC explore every interleaving of the modeled steps of the sender, the observer, the agent, and the program.
The program may exit between any 2 steps.
Both checkers give the same verdict for every configuration, and both return the shortest counterexample when a rule is removed.
The 7 properties are the 4 of Proposition~\ref{prop:safety} and 3 on the order in which the agent observes receipts.
Without its rules, the protocol violates 5 of them.
With its rules, it violates none, and a reply to a live program, an interruption, and a fresh command after a report stay reachable.
Docker tests with real tmux, real processes, and no LLM reproduce 4 of the 7 forbidden orders without the rule and pass with it, in 17 cases.
Another 13 cases test the other 3 rules, as Appendix~\ref{app:protocol} lists.

\input{fig-overview}
\input{alg-protocol}

%% file: fig-overview.tex
\definecolor{ovAcc}{RGB}{59,75,168}
\definecolor{ovHi}{RGB}{208,121,31}
\definecolor{ovInk}{RGB}{31,36,48}
\definecolor{ovMute}{RGB}{107,113,128}
\definecolor{ovLine}{RGB}{190,194,204}
\definecolor{ovBg}{RGB}{246,247,250}
\definecolor{ovTerm}{RGB}{32,36,44}
\definecolor{ovTermText}{RGB}{222,226,232}
\definecolor{ovNo}{RGB}{200,60,60}
\definecolor{ovOk}{RGB}{40,140,90}
\begin{figure}[t]
\centering
\def\ovS{\fontsize{7.2}{8.4}\selectfont}%
\def\ovF{\fontsize{6.4}{7.4}\selectfont}%
\def\ovT{\fontsize{5.6}{6.4}\selectfont}%
\begin{tikzpicture}[x=1pt,y=1pt,font=\sffamily\ovF,line width=0.5pt,>=stealth,
  band/.style={fill=ovBg,draw=ovLine,rounded corners=4pt,anchor=north west,inner sep=0pt,outer sep=0pt},
  ttl/.style={anchor=base west,inner sep=0pt,font=\sffamily\ovS\bfseries,text=ovInk},
  lbl/.style={anchor=base west,inner sep=0pt,font=\sffamily\ovT,text=ovMute},
  txt/.style={anchor=center,inner sep=0pt,font=\sffamily\ovT,text=ovInk,align=center},
  stp/.style={draw=ovAcc,fill=white,rounded corners=2.5pt,minimum height=15pt,inner xsep=3pt,inner ysep=1pt,anchor=center,font=\sffamily\ovT,text=ovInk,align=center},
  state/.style={circle,draw=#1,fill=white,minimum size=20pt,inner sep=0pt,anchor=center,font=\sffamily\ovT,text=ovInk},
  actor/.style={rectangle,rounded corners=3pt,draw=#1,fill=white,minimum size=14pt,inner sep=0pt,anchor=center,text=#1,font=\fontsize{7.5}{7.5}\selectfont},
  chip/.style={fill=ovInk,text=white,rounded corners=2pt,font=\ttfamily\ovT,inner xsep=3pt,inner ysep=2pt,anchor=center,outer sep=0pt},
  ochip/.style={chip,fill=white,draw=ovInk,text=ovInk,line width=0.4pt},
  flow/.style={->,draw=ovMute,line width=0.6pt},
  acc/.style={->,draw=ovAcc,line width=0.7pt},
  hold/.style={->,draw=ovHi,line width=0.7pt,dash pattern=on 2.2pt off 1.4pt},
  no/.style={draw=ovNo,line width=0.8pt}]

\node[band,minimum width=240pt,minimum height=42pt] at (0,0) {};
\node[ttl] at (5,-10) {1\ \ What the process waits for};
\node[ochip] (prog) at (22,-28) {migrate.sh};
\node[stp,minimum width=56pt,font=\ttfamily\ovT] (kern) at (89,-28) {read(0)\\fd 0 $\to$ tty};
\node[stp,minimum width=56pt,draw=ovAcc,fill=ovAcc!10,text=ovAcc] (det) at (168,-28) {\faCheck\ waits for input};
\draw[flow] (prog) -- (kern);
\draw[flow] (kern) -- (det);

\node[band,minimum width=240pt,minimum height=82pt] at (0,-48) {};
\node[ttl] at (5,-58) {2\ \ Order of intervention};
\node[draw=ovAcc,fill=white,rounded corners=3pt,minimum width=80pt,minimum height=27pt,anchor=north west,inner sep=0pt] (kc) at (5,-66) {};
\node[txt,font=\sffamily\ovT\bfseries,text=ovAcc] at (45,-72.5) {knowledge condition};
\node[txt,font=\rmfamily\ovF] at (45,-84) {$\mathit{deliver}(x) \Rightarrow \mathit{seen}(q)$};
\node[state=ovInk] (run) at (110,-90) {run};
\node[state=ovAcc] (wait) at (168,-90) {wait};
\node[state=ovMute] (end) at (218,-114) {\textcolor{ovMute}{end}};
\node[actor=ovInk] (agent) at (139,-116) {\faRobot};
\draw[acc] (det) -- (wait);
\draw[flow] (run) -- node[ochip,above=1.5pt,anchor=south] (ynq) {[Y/n]} (wait);
\node[anchor=base east,inner sep=0pt,font=\rmfamily\ovF,text=ovInk] at ([xshift=-1.5pt]ynq.base west) {$q$};
\draw[hold] (wait.50) .. controls (180,-62) and (196,-74) .. (wait.5);
\node[anchor=west,inner sep=0pt,font=\rmfamily\ovT,text=ovHi!85!black] (useen) at (190,-76) {$[\neg\mathit{seen}]$};
\node[chip,fill=ovHi,anchor=west] at ([xshift=2pt]useen.east) {cd ..};
\draw[acc] (wait) -- node[lbl,anchor=north west,pos=.5,inner sep=1.5pt,text=ovAcc]{\faReceipt\ receipt} (agent);
\draw[acc] (agent) -- node[chip,anchor=north east,pos=.5,xshift=-2.5pt,yshift=-2.5pt] (nch) {n} (run);
\node[anchor=east,inner sep=0pt,font=\rmfamily\ovT,text=ovAcc] at ([xshift=-2pt]nch.west) {$[\mathit{seen}]$};
\draw[flow] (wait) -- node[lbl,sloped,anchor=south,pos=.5,inner sep=1.5pt]{interrupt} (end);

\node[band,minimum width=240pt,minimum height=42pt] at (0,-136) {};
\node[ttl] at (5,-146) {3\ \ Model checking};
\node[lbl,anchor=base] at (176,-146) {\faProjectDiagram\ model};
\node[lbl,anchor=base] at (210,-146) {\faTerminal\ real};
\foreach \y/\g/\res/\c in {-157/{without $[\mathit{seen}]$}/\faTimes\ delivered/ovNo, -170/{with $[\mathit{seen}]$}/\faPause\ held/ovHi!85!black} {
  \node[lbl,anchor=west] at (5,\y) {\g};
  \node[ochip] (q\y) at (66,\y) {[Y/n]};
  \node[chip,fill=ovHi] (c\y) at (100,\y) {cd ..};
  \draw[flow] (q\y) -- (c\y);
  \draw[flow] (c\y) -- (118,\y);
  \node[lbl,anchor=west,text=\c] at (120,\y) {\res};
}
\node[text=ovNo,anchor=center,font=\ovF] at (176,-157) {\faTimes};
\node[text=ovNo,anchor=center,font=\ovF] at (210,-157) {\faTimes};
\node[text=ovOk,anchor=center,font=\ovF] at (176,-170) {\faCheck};
\node[text=ovOk,anchor=center,font=\ovF] at (210,-170) {\faCheck};

\node[fill=ovTerm,rounded corners=3pt,minimum width=240pt,minimum height=26pt,anchor=north west,inner sep=0pt] at (0,-184) {};
\node[anchor=north west,font=\ttfamily\ovF,text=ovTermText,inner sep=0pt,align=left] at (6,-189) {[rx] state=waiting-for-input program=migrate.sh on=tty input=held\\held-text=\textcolor{ovHi}{"cd .."} prompt="Recreate ...? [Y/n]"};
\node[anchor=base east,font=\sffamily\ovT,text=ovAcc!45!white,inner sep=0pt] at (235,-203.5) {\faReceipt\ receipt};
\end{tikzpicture}
\caption{\system in Terminus~2 on the migration task of Figure~\ref{fig:motivation}, where the agent has queued \cmd{cd ..} before the question appears, with the receipt that the agent reads.}\label{fig:overview}
\end{figure}
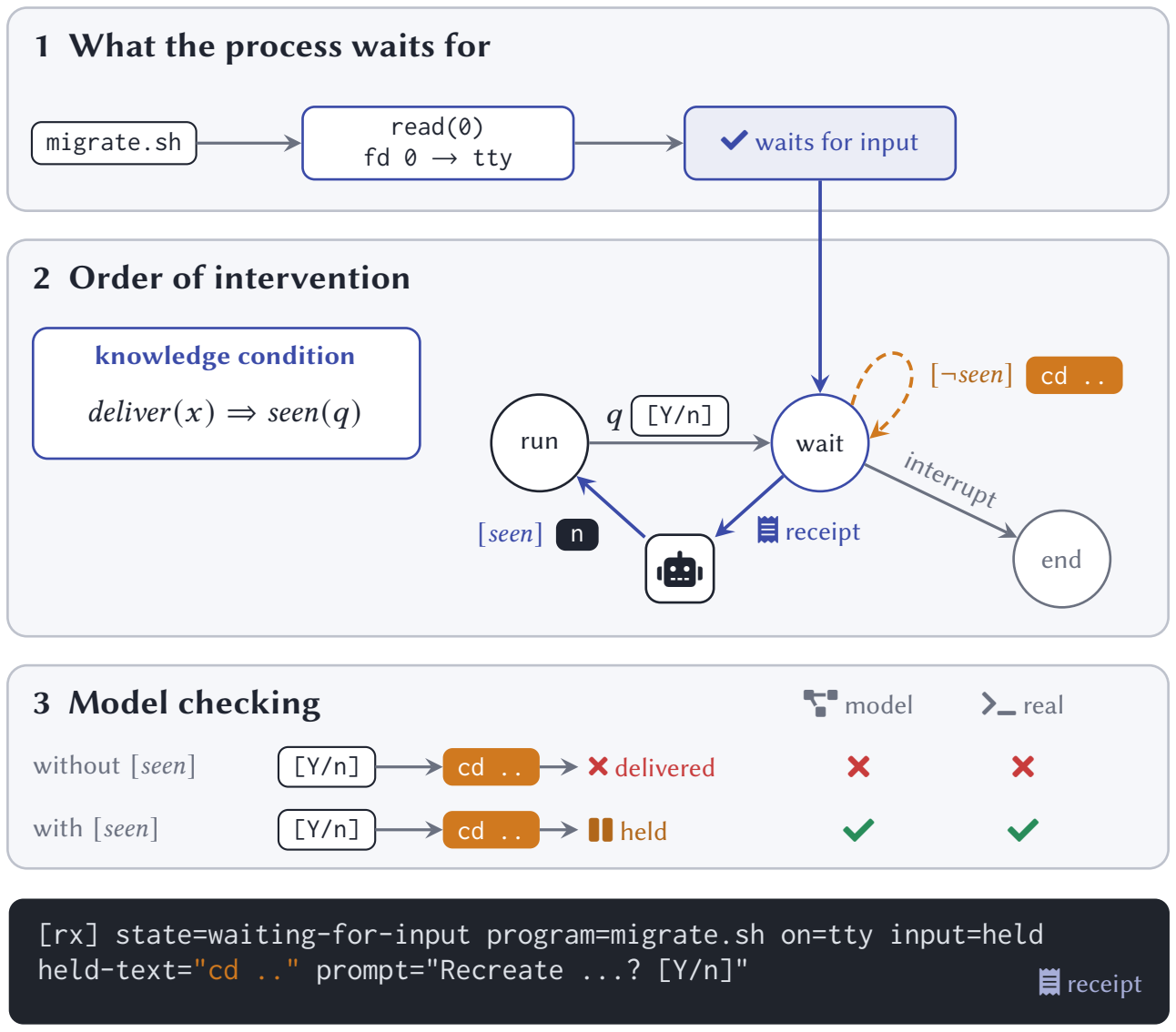

%% file: alg-protocol.tex
\begin{algorithm}[t]
\caption{The steps of the terminal protocol. The sender and the observer take each step under the lock $\ell$.}\label{alg:protocol}
\small
\begin{algorithmic}[1]
\Procedure{Send}{$x$}\Comment{next keystroke of the agent's batch}
  \If{$c < |E|$}
    \State defer $x$ and the rest of the batch
  \ElsIf{$r = \mathit{blocked} \wedge \neg \mathit{seen}$}
    \State hold $x$, append $\langle \mathit{held}, x, \mathit{prompt} \rangle$ to $E$
  \ElsIf{$r = \mathit{blocked} \wedge \neg \mathit{fits}(x, \kappa)$ for the first time}
    \State hold $x$, append $\langle \mathit{held}, x, \mathit{reason} \rangle$ to $E$
  \Else
    \State deliver $x$\Comment{$K(\sigma)$ holds}
  \EndIf
\EndProcedure
\Procedure{Observe}{$t$}\Comment{$t$ is a confirmed wait}
  \If{the batch still schedules input}
    \State \Return
  \ElsIf{$t$ took the agent's answer at its previous read}
    \State append $\langle \mathit{asked\mbox{-}again}, t \rangle$ to $E$
  \ElsIf{$s_t$ is still the waiting call}
    \State signal the tree, append $\langle \mathit{wait}, t, \mathit{survivors} \rangle$ to $E$
  \Else
    \State record a natural exit outside $E$
  \EndIf
\EndProcedure
\Procedure{Read}{}\Comment{the agent reads the next receipt}
  \State $c \gets c + 1$, and $\mathit{seen} \gets \mathit{true}$ if $E[c]$ reports $r$'s question
\EndProcedure
\end{algorithmic}
\end{algorithm}

%% file: evaluation.tex
\section{Evaluation}\label{sec:results}

\begin{itemize}[leftmargin=*,topsep=2pt,itemsep=1pt]
\item \textbf{RQ-3: What does an unanswered question cost an agent and a leaderboard, and how much does \system recover?}
We plug \system into 2 unmodified harnesses, compare each harness with and without \system on paired tasks, with 8 LLMs in mini-SWE-agent, and evaluate \system's detection on \litmus.
\end{itemize}

\begin{figure*}[t]
\centering
\includegraphics[width=0.86\linewidth]{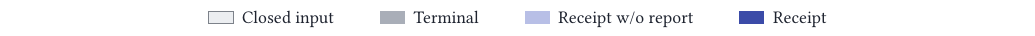}\\[-2pt]
\begin{subfigure}{0.49\linewidth}\centering
\includegraphics[width=\linewidth]{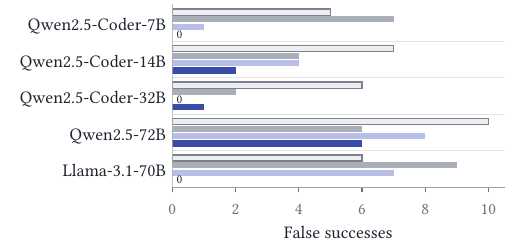}
\caption{False successes}
\end{subfigure}\hfill
\begin{subfigure}{0.49\linewidth}\centering
\includegraphics[width=\linewidth]{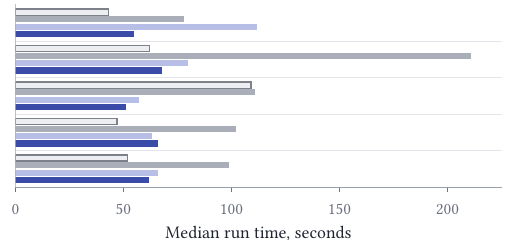}
\caption{Median run time}
\end{subfigure}
\caption{False successes and median run time of mini-SWE-agent on the \tripwire traps across 5 LLMs and 4 launch conditions.}\label{fig:agent-all}
\end{figure*}

\subsection{Benchmark}

\tripwire has 25 task pairs on database migration, deployment, asset updates, repository operations, cleanup, and credential handling.
mini-SWE-agent runs the first 15 pairs, and Terminus~2 runs all 25.
Each pair shares the task text and a \emph{verifier}, a script that checks the final state and that the agent never sees.
The 2 variants differ in a single change, such as a command option, an alias, a configuration file, or the editor that \cmd{git} opens.
This change makes the required program ask a question in the \emph{trap} variant but not in the \emph{control} variant.
In every pair, the question guards the requested action, so an unanswered question changes the final state that the verifier checks.
With input closed, the unanswered program of the first 15 pairs skips its action in 7 pairs, proceeds with a forbidden action in 4, fails in 3, and hangs in 1.
The last 5 pairs use \cmd{git rebase -i} under \cmd{nano}, a pager before \cmd{git apply}, a \cmd{whiptail} dialog, \cmd{cp -i}, and an \cmd{openssl} passphrase.
Table~\ref{tab:tasks} in Appendix~\ref{app:design} lists the changes and final-state checks of the first 15 pairs.

We plug \system into 2 unmodified harnesses and compare each harness with and without it.
Without \system, mini-SWE-agent~\citep{yang2024sweagent} 2.4.6 launches each command with stock \cmd{docker exec} and closed input or with a pseudo-terminal, and with \system it runs \system or \system w/o report.
\system w/o report is the ablation that uses the same detector and interruption rule but returns the output so far with a single-line interruption notice instead of the receipt.
Terminal and \system w/o report attach all 3 standard streams to the terminal, whereas \system attaches input to a terminal and output to a pipe.
We also repeat \system and \system w/o report with all 3 streams on the pseudo-terminal.
Terminus~2~\citep{merrill2026terminalbench} keeps a persistent terminal in a multiplexer and types the agent's keystrokes into it.
Terminus~2 runs natively without \system, or with \system or \system w/o report observing its terminal.
The 2 intervention conditions share the detector, the interruption, and the hold rule, and they differ only in the text that the agent reads.

Both harnesses use Qwen2.5-Coder-32B~\citep{hui2024qwen25coder} at temperature 0 with a limit of 30 steps per run.
mini-SWE-agent limits each command to 30 seconds and runs each variant 3 times.
Terminus~2 runs each variant once with a 600-second limit for the agent run.
We repeat mini-SWE-agent's 4 launch conditions with Qwen2.5-Coder-7B and 14B, Qwen2.5-72B~\citep{yang2024qwen25}, and Llama-3.1-70B~\citep{llama2024herd}, all Instruct versions with the same limits.
For the rankings, we add Llama-3.3-70B, Mistral-Small-24B~\citep{mistral2025small}, and Gemma-2-27B~\citep{gemma2024gemma2} with the same image and agent prompt.
We compare conditions within each task and resample the 15 tasks with their LLMs and repetitions kept together, as Appendix~\ref{app:paired} details.

\subsection{Baselines}

Beyond each harness without \system, we build 3 baselines that use the same tasks, verifiers, and budgets.
The text hint keeps terminal input and adds an input-wait message when a rule matches a prompt in the output at the 30-second limit.
The verification baseline adds ``Before you finish, verify that the task's goal state holds'' to the task, with either closed input or \system.
The post-hoc detector tests whether false successes can be found in the trajectory text after the run.
It follows Advani et al.~\citep{advani2026falsesuccess}: a TF-IDF classifier over the serialized trajectory, trained on true and false successes with the verifier as the label, and evaluated with each task pair held out.

We also use 2 published techniques as run-time baselines.
Panthi et al.~\citep{panthi2026outcome} use an outcome monitor that checks each returned tool result against contracts mined from task-disjoint traces.
When a contract fails, the monitor appends a non-binding notice that names the violated property and lists recovery options.
We mine the contracts from the shell calls of 3{,}464 resolved SWE-rebench trajectories under the cited acceptance rule and attribute a compound command to its first program.
The second baseline replaces \system's scanner with DeepSeek's process inspector, and its receipt keeps only the fields that the inspector's verdict supplies.
The inspector checks only standard input, so it does not report a program that reads \cmd{/dev/tty}.

\subsection{Metrics}

\emph{Successes} counts verified completions, and \emph{False successes} counts completion claims failing verification.
\emph{Control successes} and \emph{Trap successes} give successes over runs of each variant.
\emph{Median run time} covers setup and agent execution, excluding verification, and \emph{Tokens} sums the prompt and completion tokens of all LLM calls.
Appendix~\ref{app:design} specifies completion claims.

\emph{Waits reported} counts prompting invocations reported as input waits.
\emph{Report rate} divides this count by prompting invocations.
\emph{False positives} counts controls reported as waits.
\emph{Median report delay} summarizes the elapsed time of reported waits.

\emph{Flagged false successes} counts false successes among the runs with the highest detector scores at a given flag rate.
For input-wait prediction, \emph{AUROC} is area under the receiver operating characteristic curve, and \emph{AP} is average precision.

\subsection{Detection and Task Recovery}

The trap and the control variant of a pair differ only in the question, so their difference is the cost of the question.

Without \system, an unanswered question makes the agent misjudge the command or pay for the wait.
As Table~\ref{tab:agent-all} shows, stock mini-SWE-agent with closed input yields the most false successes across 5 LLMs.
A post-hoc trajectory classifier flags only 6 of these 34 false successes at a 10\% flag rate.
Under a plain terminal, the trap runs take 2.6 times the total run time and 2.9 times the tokens of the controls.

\begin{table}[t]
\caption{mini-SWE-agent on \tripwire traps across 5 LLMs.}\label{tab:agent-all}
\small
\setlength{\tabcolsep}{4pt}
\begin{tabular}{@{}lrrrr@{}}
\toprule
Condition & Successes & \shortstack{False\\successes} & \shortstack{Median\\run time} & \shortstack{Million\\tokens} \\
\midrule
Closed input & 153 & 34 & 53\,s & 4.4 \\
Terminal & 158 & 28 & 105\,s & 10.4 \\
\system w/o report & 162 & 20 & 66\,s & 7.8 \\
\textbf{\system} & \textbf{176} & \textbf{9} & 62\,s & 7.9 \\
\bottomrule
\end{tabular}
\end{table}

\begin{figure}[t]
\centering
\includegraphics[width=0.82\linewidth]{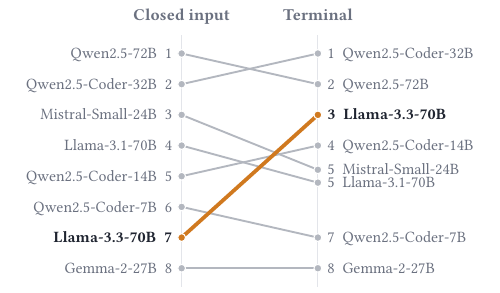}
\caption{Rank of 8 LLMs by trap successes of mini-SWE-agent on \tripwire under closed input and a plain terminal.}\label{fig:ranks}
\end{figure}

As Figure~\ref{fig:ranks} shows, the input connection also reorders the 8 LLMs by trap successes, and Kendall's $\tau_b$ between the closed-input and the terminal ranking is only 0.47.
Llama-3.3-70B ranks 7th with input closed and 3rd under a plain terminal.
Table~\ref{tab:ranks} in Appendix~\ref{app:design} lists the ranks under all 4 launch conditions.
Such waits also occur on a public benchmark, where native Terminus~2 ends 13 trials of the 38 Terminal-Bench 2.0 tasks with recorded waits at a question.

\system detects these questions from the process state.
On the 120 locally runnable Linux invocations of \litmus that wait under a terminal, its \emph{Report rate} is 85.0\%, compared with 66.7\% and 68.3\% for the text baselines, at a \emph{Median report delay} of 1.1 seconds.
The 1-second watchdog is as fast, but it ends 82 of the 119 controls, whereas \system ends none, as Appendix~\ref{app:components} details.

Plugged into mini-SWE-agent, \system lets the agent recover at a lower cost.
As Table~\ref{tab:agent-all} shows, \system raises successes from 153 with stock closed input to 176, lowers false successes from 34 to 9, and uses 24.4\% fewer tokens than a plain terminal.
As Figure~\ref{fig:agent-all} shows, it produces fewer false successes than closed input and a shorter median run than a plain terminal with every LLM.
Task-paired resampling in Appendix~\ref{app:paired} places the 95\% interval of the change in false successes below 0 against closed input, a plain terminal, and \system w/o report.
As Table~\ref{tab:matched} shows, in separate runs with identical stream connections, the receipt raises trap successes over \system w/o report from 161 to 174.
The receipt raises trap successes or lowers false successes for 4 of the 5 LLMs and worsens neither for any LLM.
As Table~\ref{tab:agent} shows, \system also has fewer false successes than the text hint, and its median trap run takes 51 seconds against 109 for the text hint.
The outcome monitor and the inspector complete fewer trap runs with more false successes.

\begin{table}[t]
\caption{mini-SWE-agent with Qwen2.5-Coder-32B on \tripwire.}\label{tab:agent}
\small
\setlength{\tabcolsep}{4pt}
\begin{tabular}{@{}lrrr@{}}
\toprule
Condition & \shortstack{Control\\successes} & \shortstack{Trap\\successes} & \shortstack{False\\successes} \\
\midrule
Closed input & 45 & 33 & 6 \\
Terminal & 45 & 40 & 2 \\
Terminal + text hint & 44 & 41 & 2 \\
\midrule
\multicolumn{4}{l}{Published baselines} \\
Outcome monitor~\citep{panthi2026outcome} & 41 & 30 & 6 \\
DeepSeek inspector~\citep{deepseek2026harness} & 40 & 35 & 3 \\
\midrule
\system w/o report & 44 & 40 & 0 \\
\textbf{\system} & 45 & 40 & 1 \\
\bottomrule
\end{tabular}
\end{table}

\begin{table}[t]
\caption{Trap successes / False successes of mini-SWE-agent on \tripwire with all 3 streams on the pseudo-terminal.}\label{tab:matched}
\small
\setlength{\tabcolsep}{4pt}
\begin{tabular}{@{}lcc@{}}
\toprule
LLM & \system w/o report & \textbf{\system} \\
\midrule
Qwen2.5-Coder-7B~\citep{hui2024qwen25coder} & 21 / 1 & \textbf{23 / 1} \\
Qwen2.5-Coder-14B~\citep{hui2024qwen25coder} & 31 / 4 & \textbf{31 / 2} \\
Qwen2.5-Coder-32B~\citep{hui2024qwen25coder} & 39 / 0 & \textbf{39 / 0} \\
Qwen2.5-72B~\citep{yang2024qwen25} & 35 / 8 & \textbf{40 / 4} \\
Llama-3.1-70B~\citep{llama2024herd} & 35 / 7 & \textbf{41 / 1} \\
\midrule
All 5 LLMs & 161 / 20 & \textbf{174 / 8} \\
\bottomrule
\end{tabular}
\end{table}

Plugged into Terminus~2, \system also removes a failure that a persistent terminal adds.
Terminus~2 returns after the agent's chosen delay, so the next scheduled keystroke becomes the answer to the question.
As Table~\ref{tab:terminus} shows, \system completes all 25 trap tasks without a false success, whereas native Terminus completes 17 and \system w/o report 20.
In the migration task of Figure~\ref{fig:motivation}, native Terminus feeds the queued \cmd{cd} command to the script's question and rebuilds the database, whereas \system reports the held text with the question and the agent answers \cmd{n}.

\begin{table}[t]
\caption{Terminus~2 with Qwen2.5-Coder-32B on \tripwire.}\label{tab:terminus}
\small
\setlength{\tabcolsep}{2.3pt}
\begin{tabular}{@{}lrrrrr@{}}
\toprule
 & \multicolumn{2}{c}{First 20 pairs} & \multicolumn{2}{c}{Last 5 pairs} & \\
\cmidrule(lr){2-3}\cmidrule(lr){4-5}
Condition & \shortstack{Trap\\successes} & \shortstack{False\\successes} & \shortstack{Trap\\successes} & \shortstack{False\\successes} & \shortstack{Control\\successes} \\
\midrule
Native & 14 & 2 & 3 & 1 & 25 \\
\system w/o report & 18 & 1 & 2 & 0 & 24 \\
\textbf{\system} & \textbf{20} & \textbf{0} & \textbf{5} & \textbf{0} & \textbf{25} \\
\bottomrule
\end{tabular}
\end{table}

\find{{\bf [RQ-3] Findings:} (1)~Without \system, an unanswered question multiplies token use by 2.9 under a plain terminal, and the input connection changes the ranking of 8 LLMs. (2)~\system reports 85.0\% of the tested waits without false positives. (3)~Plugged into mini-SWE-agent, \system reduces false successes by 73.5\% across 5 LLMs, and plugged into Terminus~2, it completes all 25 trap tasks. {\bf Insight:} A benchmark should fix and report its harness's input connection, and an agent builder should improve what the agent observes before replacing the LLM.}

%% file: deployment.tex
\section{Deployment and Lessons}\label{sec:deploy}

\system runs in 2 deployments at the developers' company.
A rule-only monitoring hook has run in a developer's Claude Code since 5 October 2026, over 11{,}694 shell calls in 41 sessions, without changing any command.
We also ported the detector to Rust for the document-review agent runtime, released on Linux and Windows on 9 October 2026 as an opt-in feature.
With the feature on, the runtime keeps each shell command's input open and records the detector's verdict for every call.
From 10 to 25 October 2026, 14 desktop installs of 12 user accounts logged 1{,}436 shell calls on Windows, and from 11 to 17 October the Web deployment's server logged 149 on Linux.
The detector's checks took a median of 0.11 milliseconds per call on Windows and 0.54 on Linux.

We draw 2 lessons for other teams that run agents against Web services.
First, the harness's input connection and time limit decide what a prompting program does and which LLM looks best.
Second, the order of input, interruption, and report needs a model checker, because each repair must not block replies or interruptions.

%% file: related.tex
\section{Related Work}\label{sec:related}

\subsection{Web Agents and Execution Interfaces}

Web-agent research measures how agents complete tasks through browser interfaces~\citep{zhou2024webarena,koh2024visualwebarena,drouin2024workarena}, workplace tasks combine Web interactions with code execution~\citep{xu2025agentcompany,chezelles2025browsergym}, and the choice of the agent's interface to the Web changes success and token use~\citep{steiner2026interfaces}.
These studies evaluate the Web's front end, whereas the agents that we study operate its back end through a shell harness.

Other work changes how an agent runs commands.
Non-blocking interfaces expose debugger and server prompts~\citep{abramovich2024enigma}, and command-text changes are studied before execution~\citep{li2026quotebench}.
Asynchronous execution lets an agent progress during tool calls~\citep{gim2024asynclm,shi2026asynctool,froger2026gaia2}.
DeepSeek's process inspector~\citep{deepseek2026harness} checks standard input and the terminal's foreground process group, whereas \system reads what each process waits for and orders its interventions under a model-checked protocol.

\subsection{Command Analysis and Task Verification}

Before a command runs, shell analyses predict its effects from system-call traces and annotations~\citep{lamprou2025caruca,vasilakis2021pash,lazarek2026sash}, repair missing non-interactive options in Dockerfile commands~\citep{durieux2023parfum}, or check its risk patterns~\citep{liu2026care}.
\system instead checks the running process, including waits with no printed prompt.

Task verification relates execution to the effects that a user requested~\citep{yang2026iec}.
State-based tests and developer-written postconditions check these effects~\citep{trivedi2024appworld,mansoor2026verified}.
Trajectory classifiers assess completion claims after the run~\citep{advani2026falsesuccess}, and contracts mined from traces supply recovery hints~\citep{panthi2026outcome}.
\system provides process-state evidence during execution and needs no task-specific postcondition.

%% file: conclusion.tex
\section{Conclusion}\label{sec:conclusion}

In this paper, we study the return assumption and present \system, which returns the process state to the agent.
Our results lead to 3 findings.
First, closing input does not make a command safe, because 11.5\% of the tested prompting invocations then return status 0 without performing the requested action.
Second, keeping input open has its own costs, because an unanswered question multiplies token use by 2.9 and a persistent terminal can deliver the agent's next keystroke as the answer.
Third, leaderboards rank input connections as well as LLMs, since the input connection moves Llama-3.3-70B from 7th to 3rd place.
Across 5 LLMs, \system reduces false successes from 34 to 9 and uses 24.4\% fewer tokens than a plain terminal.
An agent's reliability therefore depends on what its harness lets it observe as well as on its LLM.

%% file: threats.tex
\section{Threats to Validity}\label{sec:threats}

\paragraph{\textbf{Internal validity.}}
Agent runs vary, so a difference between conditions could arise by chance.
We therefore fix the temperature at 0, run each mini-SWE-agent variant 3 times, and compare conditions on the same task pairs with task-paired resampling.
The 2 intervention conditions also connect output differently.
With identical stream connections in Table~\ref{tab:matched}, the receipt still raises trap successes and lowers false successes.

\paragraph{\textbf{Construct validity.}}
An agent's completion claim or a status-0 return can misreport what a command did.
We therefore log process behavior with witnesses independent of the agent and decide task success with final-state verifiers hidden from the agent.
Manual labels could depend on the annotator.
We have 2 annotators label the production timeouts independently from a written guideline, and we report only the labels on which both agree.

\paragraph{\textbf{External validity.}}
Our findings may not hold for other harnesses and LLMs.
We therefore measure public trajectories of 23 harnesses and 19 LLMs, test launches under 8 Linux and 2 Windows harnesses, and run \tripwire with 8 LLMs from 4 families on 2 harnesses with different terminal designs.
A benchmark written together with the detector could also favor \system.
We therefore froze the detector before we wrote the last 10 of the 25 task pairs, and \system completes all 10 on Terminus~2.

\paragraph{\textbf{Ethics and safety.}}
We run every program that changes a system in a container or a scratch directory, and the \waitatlas and \system containers have no network access.
The production sessions are logs from 6 company developers who agreed to the analysis, and we do not release their text.

%% file: appendix.tex
\section{Details of the Study Design and Further Results}\label{app:design}

\begin{table*}[htbp]
\caption{Differences between the control and trap variants, and the verification target, for the first 15 \tripwire task pairs.}\label{tab:tasks}
\footnotesize
\setlength{\tabcolsep}{4pt}
\begin{tabular}{@{}>{\raggedright\arraybackslash}p{0.18\textwidth}>{\raggedright\arraybackslash}p{0.31\textwidth}>{\raggedright\arraybackslash}p{0.09\textwidth}>{\raggedright\arraybackslash}p{0.36\textwidth}@{}}
\toprule
Task and program & Control $\rightarrow$ trap & EOF outcome & Verifier target \\
\midrule
Cleanup, written script & Assume yes $\rightarrow$ request confirmation & Skip & Build cache absent \\
\addlinespace[1pt]
Amend, \cmd{git} & Editor \cmd{true} $\rightarrow$ \cmd{vim} & Hang & Clean tree, both files in HEAD, same message and commit count \\
\addlinespace[1pt]
Configure, written script & Positional profile $\rightarrow$ \cmd{--profile} or menu & Proceed & Release profile selected \\
\addlinespace[1pt]
Decrypt, written script & Passphrase file $\rightarrow$ \cmd{/dev/tty} & Fail & Expected secrets file produced \\
\addlinespace[1pt]
Deploy, written script & Positional target $\rightarrow$ \cmd{--target} or prompt & Proceed & Staging version updated, production version preserved \\
\addlinespace[1pt]
Prune, \cmd{find} & \cmd{-exec} $\rightarrow$ \cmd{-ok} & Skip & No temporary files, 3 CSV files remain \\
\addlinespace[1pt]
Stage, \cmd{git add} & Whole-file staging $\rightarrow$ \cmd{-p} & Skip & Required source changes staged, no new commit \\
\addlinespace[1pt]
Clean, \cmd{git clean} & \cmd{-df} $\rightarrow$ \cmd{-di} & Skip & Clean working tree, build directory retained \\
\addlinespace[1pt]
Migrate, written script & Extra column triggers a rebuild prompt & Proceed & New column present, the 3 original rows unchanged \\
\addlinespace[1pt]
Purge, \cmd{rm} & Alias \cmd{rm -f} $\rightarrow$ \cmd{rm -i} & Skip & No rotated logs, both active log files exist \\
\addlinespace[1pt]
Protect key, \cmd{ssh-keygen} & Supplied passphrase $\rightarrow$ omit \cmd{-N} & Proceed & Key accepts supplied passphrase, rejects empty passphrase \\
\addlinespace[1pt]
Rotate key, \cmd{ssh-keygen} & Old key absent $\rightarrow$ overwrite prompt & Fail & Matching new key pair with a changed fingerprint \\
\addlinespace[1pt]
Extract, \cmd{tar} & \cmd{-xzf} $\rightarrow$ \cmd{-xwzf} & Skip & Expected version and helper file present \\
\addlinespace[1pt]
Update assets, \cmd{unzip} & Destination files absent $\rightarrow$ existing files & Fail & New version markers in script and style, index exists \\
\addlinespace[1pt]
Select tool, \cmd{update-alternatives} & \cmd{--set} $\rightarrow$ \cmd{--config} & Skip & Selected executable reports the required version \\
\bottomrule
\end{tabular}
\end{table*}

\begin{table*}[htbp]
\caption{Outcome of the \litmus prompting invocations per category and input condition.}\label{tab:category}
\footnotesize
\setlength{\tabcolsep}{4pt}
\begin{tabular}{@{}l*{15}{r}@{}}
\toprule
 & \multicolumn{5}{c}{Closed} & \multicolumn{5}{c}{Pipe} & \multicolumn{5}{c}{Terminal} \\
\cmidrule(lr){2-6}\cmidrule(lr){7-11}\cmidrule(l){12-16}
Category & Hang & Fail & Skip & Unasked & Unaffected & Hang & Fail & Skip & Unasked & Unaffected & Hang & Fail & Skip & Unasked & Unaffected \\
\midrule
Confirmation & 2 & 29 & 10 & 5 & 1 & 34 & 8 & 0 & 4 & 1 & 30 & 0 & 0 & 0 & 1 \\
Credential & 4 & 26 & 0 & 2 & 0 & 20 & 13 & 0 & 0 & 0 & 28 & 0 & 0 & 0 & 0 \\
Interpreter & 1 & 2 & 1 & 0 & 17 & 19 & 2 & 0 & 0 & 0 & 14 & 0 & 0 & 0 & 0 \\
Editor & 5 & 9 & 2 & 1 & 0 & 7 & 8 & 1 & 1 & 0 & 15 & 0 & 0 & 0 & 0 \\
Pager & 0 & 0 & 0 & 0 & 13 & 1 & 0 & 0 & 0 & 12 & 10 & 0 & 0 & 0 & 1 \\
Wizard & 2 & 8 & 0 & 3 & 0 & 8 & 3 & 0 & 2 & 0 & 13 & 0 & 0 & 0 & 0 \\
Menu & 0 & 7 & 5 & 1 & 0 & 12 & 1 & 0 & 0 & 0 & 13 & 0 & 0 & 0 & 0 \\
\bottomrule
\end{tabular}
\end{table*}

\subsection{Corpus and Scoring Details}

The 39 Terminal-Bench 2.0 submissions that record each shell call cover 23 harnesses in 16{,}063 trials.
Of their 557{,}909 shell calls, 486{,}879 run commands and 456{,}085 have observed outcomes.
The 37 single-LLM submissions among them cover 19 LLMs.
In these submissions, a multiplexer command is text sent to a shell prompt and submitted with Enter.
Public-command replays restore the preceding commands before changing the waiting command's input.
Each input-witness run has an 8-second limit.
On Windows, we use Claude Code 2.1.286 with Git Bash and headless Codex CLI 0.153.4 with PowerShell.

\begin{table}[htbp]
\caption{Lifetime matrix under Codex CLI~\citep{openai2026codex} 0.153.4 in headless mode.}\label{tab:codex}
\small
\begin{tabular}{@{}lll@{}}
\toprule
Launch & True outcome & Agent's answer \\
\midrule
foreground & \textbf{killed at 10-14\,s of 100\,s} & \textbf{``still running''} \\
\cmd{Start-Job} & did not start & ``unknown'' \\
\cmd{Start-Process} & outlives the session, completes & ``unknown'' \\
\cmd{cmd start /b} & outlives the session, completes & ``still running'' \\
\cmd{bash nohup \&} & outlives the session, completes & ``still running'' \\
\bottomrule
\end{tabular}
\end{table}

\begin{table}[htbp]
\caption{Lifetime matrix under Claude Code~\citep{anthropic2026claudecode} 2.1.286.}\label{tab:lifetime}
\small
\begin{tabular}{@{}llll@{}}
\toprule
Launcher & Launch & Outcome & Notice \\
\midrule
main & harness background & completes & yes \\
main & foreground past limit & background, completes & yes \\
main & shell background & completes & no \\
sub-agent & harness background & \textbf{killed at 2-6\,s of 150\,s} & \textbf{no} \\
sub-agent & \cmd{nohup} or \cmd{\&} & completes & no \\
\bottomrule
\end{tabular}
\end{table}

In the Linux launch comparison, each harness talks to a scripted LLM server.
In Ubuntu 24.04, we test Claude Code 2.1.292, Codex CLI 0.160.1, Gemini CLI 0.63.0, OpenCode 1.18.35, mini-SWE-agent 2.4.6, Aider 0.86.2, Goose 1.30.0, and Terminus~2 from Harbor 0.24.0.
We also pass 12 prompting invocations through each harness.

Claude Code's Bash tool closes input for a plain command but supplies an open pipe when the command contains a here-document.
In the production sessions, 18.5\% of the 34{,}891 Bash calls contain a here-document.

Completion scoring distinguishes a failure report from an unsupported success claim.
In every mini-SWE-agent condition, the task requests a final \cmd{STATUS: DONE} or \cmd{STATUS: NOT DONE}.
We use the last status and count a submission without a status as a completion claim.
An error or step-limit termination without submission is not a claim.
A submitted \cmd{STATUS: NOT DONE} is not a false success even when verification fails.
For Terminus~2, a completion claim requires \texttt{mark\_\allowbreak task\_\allowbreak complete} in the final step after an earlier completion call.
A Terminus~2 success requires the target final state, whether or not the named script runs.

\begin{table}[htbp]
\caption{Successes / False successes / Median run time in seconds of mini-SWE-agent on the 15 \tripwire traps.}\label{tab:models}
\footnotesize
\setlength{\tabcolsep}{1.5pt}
\begin{tabular}{@{}lcccc@{}}
\toprule
LLM & Closed & Terminal & w/o report & \textbf{\system} \\
\midrule
Qwen2.5-Coder-7B~\citep{hui2024qwen25coder} & 26 / 5 / 43 & 25 / 7 / 78 & 21 / 1 / 112 & \textbf{26 / 0 / 55} \\
Qwen2.5-Coder-14B~\citep{hui2024qwen25coder} & 29 / 7 / 62 & 31 / 4 / 211 & 31 / 4 / 80 & \textbf{32 / 2 / 68} \\
Qwen2.5-Coder-32B~\citep{hui2024qwen25coder} & 33 / 6 / 109 & 40 / 2 / 111 & 40 / 0 / 57 & \textbf{40 / 1 / 51} \\
Qwen2.5-72B~\citep{yang2024qwen25} & 35 / 10 / 47 & 33 / 6 / 102 & 35 / 8 / 63 & \textbf{38 / 6 / 66} \\
Llama-3.1-70B~\citep{llama2024herd} & 30 / 6 / 52 & 29 / 9 / 99 & 35 / 7 / 66 & \textbf{40 / 0 / 62} \\
\midrule
All 5 LLMs & 153 / 34 / 53 & 158 / 28 / 105 & 162 / 20 / 66 & \textbf{176 / 9 / 62} \\
\midrule
Llama-3.3-70B~\citep{llama2024herd} & 25 / 14 / 39 & 32 / 4 / 81 & 34 / 4 / 49 & \textbf{32 / 7 / 51} \\
Mistral-Small-24B~\citep{mistral2025small} & 32 / 6 / 47 & 29 / 5 / 121 & 37 / 1 / 46 & \textbf{37 / 2 / 53} \\
Gemma-2-27B~\citep{gemma2024gemma2} & 9 / 13 / 22 & 11 / 2 / 183 & 31 / 1 / 26 & \textbf{25 / 0 / 33} \\
\bottomrule
\end{tabular}
\end{table}

The 3 LLMs below the total of Table~\ref{tab:models} ran with the same image and agent prompt.
These LLMs enter only the rankings of Table~\ref{tab:ranks}.
Gemma-2-27B ran on vLLM 0.10.1 with the system message at the start of the first user turn.

\begin{table}[htbp]
\caption{Rank of each LLM by trap successes of mini-SWE-agent on 15 \tripwire task pairs under each launch condition.}\label{tab:ranks}
\small
\setlength{\tabcolsep}{3pt}
\begin{tabular}{@{}lcccc@{}}
\toprule
LLM & Closed & Terminal & w/o report & \textbf{\system} \\
\midrule
Qwen2.5-72B~\citep{yang2024qwen25} & 1 & 2 & 3 & \textbf{3} \\
Mistral-Small-24B~\citep{mistral2025small} & 3 & 5 & 2 & \textbf{4} \\
Qwen2.5-Coder-32B~\citep{hui2024qwen25coder} & 2 & 1 & 1 & \textbf{1} \\
Llama-3.1-70B~\citep{llama2024herd} & 4 & 5 & 3 & \textbf{1} \\
Qwen2.5-Coder-14B~\citep{hui2024qwen25coder} & 5 & 4 & 6 & \textbf{5} \\
Qwen2.5-Coder-7B~\citep{hui2024qwen25coder} & 6 & 7 & 8 & \textbf{7} \\
Llama-3.3-70B~\citep{llama2024herd} & 7 & 3 & 5 & \textbf{5} \\
Gemma-2-27B~\citep{gemma2024gemma2} & 8 & 8 & 6 & \textbf{8} \\
\bottomrule
\end{tabular}
\end{table}

\subsection{Detector Components Under a Common Launch}\label{app:components}

\begin{table}[htbp]
\caption{Waits reported and False positives under \system on the Linux invocations of \litmus.}\label{tab:rx}
\small
\begin{tabular}{@{}lrrr@{}}
\toprule
Outcome with input closed & $n$ & Pipe & Terminal \\
\midrule
Skip & 14 & 13 & 14 \\
Unasked & 10 & 4 & 5 \\
Hang & 7 & 7 & 7 \\
Fail & 68 & 51 & 63 \\
Unaffected & 20 & 13 & 12 \\
No outcome with input closed & 1 & 1 & 1 \\
\midrule
All prompting invocations & 120 & 89 & 102 \\
Controls: False positives & 119 & 1 & 0 \\
\bottomrule
\end{tabular}
\end{table}

The text baselines are a prompt-pattern rule and a classifier using short character sequences, both reading the last 160 characters of a tool result.
The inactivity watchdog ends a command after 1, 3, or 5 seconds without output, run 2 times per threshold under a terminal with an 8-second hard cap.
The 119 controls include long computations, sleeps, network waits, and pagers that print and exit.
Of the 18 waits that \system does not report under a terminal, 15 exit or act without waiting under \system's launch, and 3 are \cmd{npm} commands that draw a spinner while they wait.

We compare DeepSeek's inspector at commit \cmd{d743267} with \system's scanner on 120 prompting invocations and 119 controls.
Both components inspect each invocation at 1, 3, and 7 seconds in 2 repetitions, with an 8-second cap and no interruption upon detection.
\system requires input-wait system calls, and DeepSeek checks foreground process groups and stdin waits.
At 7 seconds, \system alone confirms 37 invocations, and DeepSeek alone reports 1.

\subsection{Baseline Details}\label{app:baselines}

Both baselines used the tasks and limits of Table~\ref{tab:agent}.
With the verification instruction, closed input completes 35 of the 45 trap runs with 5 false successes, and \system completes 39 with 1 false success.
With strict attribution of a result to a single program, the contracts cover 97 of the 2{,}709 commands in the 450 closed-input runs of the 5 LLMs and produce no notice.
With attribution to the first program, they produce a notice in 68 of the 225 trap runs and 57 of the 225 control runs, and in 8 of the 34 false successes.

\subsection{Native Windows Backend}\label{app:windows-receipt}

The Windows backend applies the same wait-confirmation rule through a Job Object and pending \cmd{NtReadFile} calls on the controlled pipe.
On x64 Windows 11, 26 native cases with cmd, Windows PowerShell, Python, and Node.js give the expected outcome in every case.

\subsection{Protocol Model and Tests}\label{app:protocol}

The automaton of Section~\ref{sec:repair} has a sender, an observer, a single lock, a cursor that the agent advances when it reads a receipt, and a single program per epoch that may block, exit, or survive a signal.
The sender and the observer take the lock for a check and the action that follows it, and the program may exit between any 2 steps.
When the program exits between the observer's confirmation and the observer's signal, the recheck finds the reader gone, sends no signal, and records a natural exit outside the receipt log.
The model treats the recheck and the signal as a single step and starts each command with a single blocked reader.
TLC checks the 7 properties as invariants of the TLA+ specification over 499 states.
Without its rules, the protocol violates 5 of the 7 properties, and with the first 5 rules it violates none.
A second TLA+ module checks the last 2 rules on a single reader that asks 4 questions.
Its 2 invariants forbid an interruption at a question whose report the agent has not read and the delivery of a mismatched input before the agent has read the reason.
The protocol without these rules violates the invariants in 145 states, and the final protocol satisfies them in all 114 reachable states.
Table~\ref{tab:protocol} lists the 7 forbidden orders, the rule for each, and its Docker test with real tmux and processes.

\begin{table}[htbp]
\caption{The 7 orders the terminal protocol forbids, the rule for each, and its test.}\label{tab:protocol}
\footnotesize
\setlength{\tabcolsep}{3pt}
\begin{tabular}{@{}p{0.27\columnwidth}p{0.34\columnwidth}p{0.31\columnwidth}@{}}
\toprule
Forbidden order & Rule & Test \\
\midrule
Stale input after an interruption reaches the shell & Sender and observer share a single lock, and the sender defers input published after the agent's last observation & \cmd{input\_guard}, 3 cases \\
A scheduled input is interrupted before delivery & The observer pauses while the batch still schedules input & \cmd{queued\_input}, 2 cases \\
A receipt reports a wait of a program that exited by itself & The observer rechecks the reader's system call under the lock before it signals & \cmd{exit\_race}, 8 cases \\
A scheduled keystroke answers a prompt the agent has not seen & The sender holds the keystroke, records it with the prompt, and the program keeps waiting & \cmd{hold\_on\_prompt}, 4 cases \\
A receipt published after the observation defers the agent's next batch & The observation of a batch waits for a pending candidate wait, at most 1.2\,s & \cmd{await\_receipt}, 4 cases \\
A keystroke whose form does not fit the reader's class is delivered & The sender holds it once with the reason and delivers it when sent again & \cmd{state\_receipt}, 6 cases \\
A reader the agent has just answered is interrupted & The observer reports it again and interrupts only after the agent reads that report & \cmd{state\_receipt}, 3 cases \\
\bottomrule
\end{tabular}
\end{table}

\subsection{Returned Observation Under Matched Streams}\label{app:matched}

With all 3 standard streams on the same pseudo-terminal and the same detector, interruption rule, task pairs, and budgets, \system w/o report and \system differ in the returned text.
With Qwen2.5-Coder-32B over 3 repetitions, both complete all 45 controls with no false success, and their median trap runs take 46 seconds each.
For the other 4 LLMs of Table~\ref{tab:matched}, the receipt condition uses the same image and agent prompt as their \system w/o report runs of Table~\ref{tab:models}.

\subsection{Task-Paired False-Success Analysis}\label{app:paired}

The paired analysis averages each task's false-success difference over its 5 LLMs and 3 repetitions.
\emph{Change in pp} averages these task-level changes in false-success frequency from a baseline to \system, in percentage points.
\emph{Fewer}, \emph{More}, and \emph{Tied} count tasks by the direction of this change.
Percentile 95\% intervals use 20{,}000 resamples of the 15 tasks with seed 20261009.

\begin{table}[htbp]
\caption{Task-paired changes in false successes with \system on 15 \tripwire task pairs, with 95\% intervals.}\label{tab:paired}
\small
\setlength{\tabcolsep}{3pt}
\begin{tabular}{@{}lrrrl@{}}
\toprule
Compared with & Fewer & More & Tied & Change in pp \\
\midrule
Closed input & 8 & 1 & 6 & $-11.1$ [$-20.9$, $-3.1$] \\
Terminal & 8 & 1 & 6 & $-8.4$ [$-13.8$, $-3.6$] \\
\system w/o report & 7 & 2 & 6 & $-4.9$ [$-11.1$, $-0.4$] \\
\bottomrule
\end{tabular}
\end{table}

\subsection{Prediction Before Launch}\label{app:prediction-details}

The 267 test commands come from 110 program families absent from training and include 14 pipe waits and 32 terminal waits.
Learned predictors use command text with up to 400 tokens of help text.
We tune Qwen3-0.6B\footnote{\url{https://huggingface.co/Qwen/Qwen3-0.6B}} for 2 passes at learning rate $2\times10^{-5}$ with 5 seeds.
Zero-shot Qwen2.5-32B-Instruct~\citep{yang2024qwen25} uses greedy decoding and a 64-token limit.

\begin{table}[htbp]
\caption{Wait prediction on held-out program families of \waitatlas.}\label{tab:predict}
\small
\begin{tabular}{@{}lrrrr@{}}
\toprule
 & \multicolumn{2}{c}{Pipe} & \multicolumn{2}{c}{Terminal} \\
Predictor & AUROC & AP & AUROC & AP \\
\midrule
Rule set & 0.603 & 0.170 & 0.572 & 0.199 \\
Logistic regression & 0.737 & 0.126 & 0.704 & 0.222 \\
Tuned LLM & 0.761 & 0.194 & 0.666 & 0.240 \\
Zero-shot 32B & 0.900 & 0.477 & 0.717 & 0.395 \\
\bottomrule
\end{tabular}
\end{table}

As Table~\ref{tab:predict} shows, zero-shot prediction ranks waits above returned commands on held-out families.
On the 434 public waits of Terminal-Bench 2.0, the rule set finds 25.3\%, and it finds none of the 178 waits in programs outside \waitatlas.